\documentclass[conference]{IEEEtran}
\IEEEoverridecommandlockouts
\usepackage{cite}
\usepackage{amsmath,amssymb,amsfonts}
\usepackage{graphicx}
\usepackage{textcomp}
\usepackage{xcolor}
\usepackage{multicol}			% multiple columns in the text
\usepackage{multirow}			% mutiple rows in the text
\usepackage{pdfpages}
\usepackage{subfig}
\usepackage{float}
\usepackage{algorithm}
\usepackage{algpseudocode}
\usepackage{tikz}
\usepackage{ifpdf}
\input{Qcircuit}

\newcommand\copyrighttext{%
  \footnotesize \copyright 2020 IEEE.
Personal use of this material is permitted. 
Permission from IEEE must be obtained for all other uses, in any current or future media, including\\	 reprinting/republishing this material for advertising or promotional purposes,creating new collective works, for resale or redistribution to servers or lists, or reuse of any copyrighted component of this work in other works.}
\newcommand\copyrightnotice{%
\begin{tikzpicture}[remember picture,overlay]
\node[anchor=south,yshift=10pt] at (current page.south) {\fbox{\parbox{\dimexpr\textwidth-\fboxsep-\fboxrule\relax}{\copyrighttext}}};
\end{tikzpicture}%
}

\def\BibTeX{{\rm B\kern-.05em{\sc i\kern-.025em b}\kern-.08em
    T\kern-.1667em\lower.7ex\hbox{E}\kern-.125emX}}
\begin{document}
%% Copyright notice

\title{Quantum ensemble of trained classifiers}

\author{
\IEEEauthorblockN{Ismael C. S. Araujo}
\IEEEauthorblockA{\textit{Departamento de Computação} \\
\textit{Universidade Federal Rural de Pernambuco}\\
Recife, Pernambuco, Brazil \\
ismael.cesar@ufrpe.br}
\and
\IEEEauthorblockN{Adenilton J. da Silva}
\IEEEauthorblockA{\textit{Centro de informática } \\
\textit{Universidade Federal de Pernambuco}\\
Recife, Pernambuco, Brazil\\
ajsilva@cin.ufpe.br}
}

\maketitle

\begin{abstract}

Through superposition, a quantum computer is capable of representing an exponentially large set of states, according to the number of qubits available. 
Quantum machine learning is a subfield of quantum computing that explores the potential of quantum computing to enhance machine learning algorithms.
An approach of quantum machine learning named quantum ensembles of quantum classifiers consists of using superposition to build an exponentially large ensemble of classifiers to be trained with an optimization-free learning algorithm.
In this work, we investigate how the quantum ensemble works with the addition of an optimization method. Experiments using benchmark datasets show the improvements obtained with the addition of the optimization step.
\end{abstract}

% Sections -------------------------------------------------------------------------
\copyrightnotice

\section{Introduction}

Artificial intelligence (AI) is a field of computer science that studies the creation of programs designed to act as intelligent agents, created to evaluate and automatically make decisions based on the inputs~\cite{russell2016artificial}.
Machine learning (ML) is a subfield of AI that studies the creation of algorithms and programs that are capable of not only make decisions based on the inputs but also learn with it to improve performance~\cite{faceli2011inteligencia, mitchell1997machine, bishop2006pattern}.
In this work, we name the programs and algorithms that involve some ML method as ML models, or only models.

Quantum computing is a field of computer science that studies the codification and processing of information in quantum systems\cite{nielsen2000quantum}.
The smallest unit that represents information in a quantum computer is a quantum bit or \textit{qubit}.
Unlike a classical bit, which can only assume one state at a time, either $0$ or $1$ exclusively, a \textit{qubit} has the property of being in both states $0$ and $1$ at the same time. 
This superposition of information means that a quantum computer with multiple \textit{qubits} is capable of representing an exponentially large number of states, according to the number of qubits available.

ML models usually have to process feature vectors with large dimensions, which can impact the cost of processing time and memory consumption.
The power of state representation makes quantum computing a candidate for the improvement of ML models.
Quantum Machine Learning (QML) is the subfield of quantum computing where researchers have been studying how to explore the potential of quantum computing to enhance ML~\cite{biamonte2017quantum, 2015introduction}.

In \cite{schuld2018quantum}, a QML approach named quantum ensemble of quantum classifiers is proposed. 
Classifier outputs weigh the degree of influence of each classifier in the final answer. 
In \cite{schuld2018quantum}, the accuracy of each classifier weighs its significance, and bad performing classifiers would have a small impact on the final output.

In a quantum version of this kind of ensemble, it would be possible to use superposition to create an exponentially large set of classifiers, with the accuracy codified in the probability amplitude of the state. 
We could use this quantum ensemble as an optimization free model, 
with the supposition that an exponentially large ensemble of classifiers weighed by its accuracy would return good classifications~\cite{schuld2018quantum}. 

In this work, we explore the idea of using quantum computers to efficiently create an ensemble of classifiers and perform experiments based on simulations using benchmark datasets.
We propose a quantum ensemble strategy based on \cite{schuld2018quantum}  with a training phase.
We show the differences between quantum ensembles with optimized and unoptimized classifiers.
The ML model used in the simulations was ANNs, so as for the therms optimized and unoptimized were taken as trained and untrained ANNs.

The rest of this work is organized as follows: 
Section \ref{sec:quantum_computing} contains some introductory explanations concerning quantum computing.
Section \ref{sec:quantum_ensembles} introduces a more detailed description of quantum ensembles mentioned in the introduction, as well as replications of numerical analysis made in \cite{schuld2018quantum}.
Section \ref{sec:methodology} presents the main contribution, describes the simulations, presents metrics used to evaluate the model, and shows the results concerning the presented methodology. 
Finally, Section \ref{sec:conclusions} contains some remarks about the results presented, as well as possible future works.

\section{Quantum computing}
\label{sec:quantum_computing}
% Single qubit states
The base unit of information used in a quantum computer is called a quantum bit or \textit{qubit}.
Mathematically a \textit{qubit} can be represented in the form of a column vector, however, for simplicity and convenience, many works in the literature rather use the Dirac's notation~\cite{nielsen2000quantum, mcmahon2007quantum}. 
The symbol $\ket{\cdot}$ is called ``ket'' and the equivalence between Dirac's notation and column vector notation is made in the following equation:

\begin{equation}
    \begin{array}{c}
    \left[\begin{array}{c c}    1 & 0 \\ \end{array} \right]^T = \ket{0} \\ \\
    \left[\begin{array}{c c}    0 & 1 \\ \end{array} \right]^T = \ket{1}
    \end{array}
\end{equation}

A \textit{qubit} has the property to be in a state of superposition, that is, to assume more than one state at the same time. 
The superposition can be mathematically represented by a linear combination of the basis states.
The coefficients in the linear combination would represent the amplitude probability of the system state to assume a specific state.
More specifically, Let the state $\ket{\psi}$ in Eq. \eqref{eq:sigle_qubit_superposition} be an arbitrary state of a single \textit{qubit}.

The coefficients $\alpha$ and $\beta$ are complex numbers representing the amplitude probability of $\ket{\psi}$.

\begin{equation}
    \ket{\psi} = \alpha \ket{0} + \beta \ket{1}
    \label{eq:sigle_qubit_superposition}
\end{equation}

% Multiqubit state

A quantum state can also be composed of multiple \textit{qubits}.

We use tensor products to represent multiple qubits. 
For instance, the two-\textit{qubit} state $\ket{0} \otimes \ket{1} = \ket{01}$. 
Similarly to a single \textit{qubit} state, the \textit{multiqubit} state can be in superposition.
Let $x \in \{0,1\}^n$ be a binary string, where $n$ is the number of \textit{qubits} in the string, and $\ket{\psi'}$ a \textit{multiqubit} in superposition.
Eq.~\eqref{eq:multiqubit_superpoistion} describes the \textit{multiqubit} $\ket{\psi}$.

\begin{equation}
    \ket{\psi} = \sum_{x \in \{0,1\}^n} \alpha_{x}\ket{x}
    \label{eq:multiqubit_superpoistion}
\end{equation}

% measurement
To extract useful information from a superposed state, we need to perform a measurement operation. 
However, it is impossible to measure each component of a superposed state without affecting the entire system. 
That is because when we apply a measurement operation is applied to the system, the superposition of the system collapses, and the system assumes a specific state.
The information contained in other components of the superposition is lost or changed. 
For example, if the qubit $\ket{\psi}$ in Eq. \eqref{eq:sigle_qubit_superposition} is measured, $\ket{\psi}$ will collapse to  the state $\ket{0}$  with $|\alpha|^2$ probability and $\ket{1}$ with $|\beta|^2$ probability,
where $|\alpha|^2 + |\beta|^2 = 1$. 

Another case to be analyzed is when a \textit{multiqubit} system is partially measured.
Suppose the two last qubits of the state in Eq.\eqref{eq:multiqubit_superpoistion} are being measured.
And let $\bar{x} \in \{\{0,1\}^{n-2} + 01\}$ be the regular expression that represents the binary strings of size $n$ that end with $01$. 
After the measurement of the two last qubits is performed, the system would return $\ket{01}$ with a probability $p = \sum_{\bar{x}} |\alpha_{\bar{x}}|^2$. 
And the state $\ket{\psi'}$ would collapse to the state $\ket{\psi''}$, as shown in Eq.\eqref{eq:multiqubit_state_partial_measurement}

\begin{equation}
    \ket{\psi''} = \sum_{\bar{x} \in \{\{0,1\}^{n-2} + 01\}} \frac{ \alpha_{\bar{x}} \ket{\bar{x}} }{ \sqrt{ \sum_{\bar{x}} |\alpha_{\bar{x}}|^2 } }
    \label{eq:multiqubit_state_partial_measurement}
\end{equation}

% Quantum gates
Quantum operators, or quantum gates, are used to operate and manipulate the states and amplitude probabilities of quantum systems.
Let $N = 2^n$ be the number of states that can be represented with $n$ \textit{qubits}.
Given a fixed basis, an $\mathbb{C}^{N\times N}$ unitary matrix represents a quantum gate over $n$ qubits.
Quantum gates can also be combined using tensor products to perform multiqubit operations. For instance, $I\otimes X\otimes I$,
where $I$ is the single-\textit{qubit} identity gate and $X$ the \textit{not} gate.

Since quantum gates work as linear operators, these operators can modify different components of a superposition linearly, this property is known as quantum parallelism \cite{nielsen2000quantum, mcmahon2007quantum, yanofsky2008quantum}.
In Equation~\eqref{eq:apply_ixi} an application of the gates $I \otimes X \otimes I$ on a superposed three-\textit{qubit} state is exemplified.

\begin{equation}
    \begin{array}{c}
\left(I \otimes X \otimes I\right) \left(\gamma\ket{010}+\phi\ket{110}\right) \rightarrow  \gamma\ket{000}+\phi\ket{100}
    \end{array}
    \label{eq:apply_ixi}
\end{equation}

% Quantum Parallelism and quantum machine learning
A quantum computer can theoretically simulate any binary function of a classical computer~\cite{nielsen2000quantum}.
Let $A$ be a quantum operator that implements a classical function $f(x)$, such that $A\ket{x}\ket{0} = \ket{x}\ket{f(x)}$. 
Eq.~\eqref{eq:apply_A_to_superposition} shows an example of the action of the $A$ operator in a superposed state.

$A$ is a linear operator and can be applied to each component of the superposition thanks to quantum parallelism. 
With superposition and quantum parallelism, it is possible to apply a function in an exponentially large number of states at the same time.

\begin{equation}
    \sum_{x \in \{0,1\}^n}  \alpha_{x}A\ket{x}\ket{0} = \sum_{x \in \{0,1\}^n}  \alpha_{x}\ket{x}\ket{f(x)}
    \label{eq:apply_A_to_superposition}
\end{equation}

\section{Quantum Ensembles}
\label{sec:quantum_ensembles}
A classifier can be described as a function that maps features into the classes: $f:\mathcal{X} \rightarrow \mathcal{Y}$, where $\mathcal{X}$ is the space of features and $\mathcal{Y}$ is the space of classes.
A parameterized classifier makes the mapping of features according to the parameters that were set, such as the mapping $f: \mathcal{X} \times \Theta \rightarrow \mathcal{Y}$ where $\Theta$ is the space of the parameters.
However in this work a parameter $\theta \in \Theta$ shall be referred to as the classifier itself. 
Given that in order to improve a classification model's performance, one can optimize its parameters $\theta$, which is the case with artificial neural networks (ANNs)~\cite{bishop2006pattern, haykin2001neural}.

Thus, an ensemble of classifiers is a classification method in which the output of different classifiers is combined to obtain a final answer. 
For example, given a binary classification problem where the set of classes is $Y = \{-1, 1\}$.
Let the set of classifiers be $E = \{\theta_{0}, \theta_{1}, \ldots, \theta_{n-1}\}$. 
Given an unseen data sample $\tilde{x}$, the output of each classifier is combined in a sum in order to obtain the ensemble's final answer as it is shown in Eq. \eqref{eq:basic_classical_ensemble}.

\begin{equation}
    \tilde{y} = sign \left( \sum_{\theta \in E} w_{\theta} f(\tilde{x}, \theta)  \right)
    \label{eq:basic_classical_ensemble}
\end{equation}

Where $sign$ is the sign function and  $\tilde{y}$ is the answer (class) returned by the ensemble.
Where the output of each classifier is pondered by a factor $w_{\theta}$, which represents its desired degree of influence in the sum. 
A way to determine the value of the factor $w_{\theta}$ is by using the classifier's accuracy, which can be obtained by measuring the performance of each classifier in the dataset before using the ensemble to label new data~\cite{schuld2018quantum}.
This way, bad performing classifiers would have little influence in the sum. 

\subsection{Quantum ensemble described in quantum states}
In order to implement a quantum version of an ensemble such as in Eq. \eqref{eq:basic_classical_ensemble}, the system would have to be divided into five quantum registers. 
Taking into consideration the data would be encoded in the qubits, the first register would be dedicated do store the data $\ket{x}$. 
The data can be processed in the system sample by sample, or it can be stored in superposition using a storing procedure from some kind probabilistic, or associative quantum memory~\cite{ventura1998artificial, ventura1999quantum, trugenberger2002quantum}.
For simplicity, a sample by sample approach will be considered.

The second register would be used for storing the classifiers $\ket{\theta}$. 
The third and fourth register would be used for storing the answer of each classifier $\ket{\hat{y}_\theta}$ and the correct answer $\ket{y}$ respectively.
And the fifth register would be a one \textit{qubit} register used as ancilla.

For the initial state of the ensemble, all the classifiers would have to be stored in superposition. 
And the ancillary register would be initialized in superposition as well.
Let $x$ be one data sample from a dataset $\mathcal{D}$ with its label being $y$. 
Both the data sample and its label would be stored in their respective registers and the answer register $\ket{\hat{y}_\theta}$ would be initialized in the state $\ket{0}$. 
Therefore, the initial state for the ensemble would be as in Eq. \eqref{eq:q_ensemble_intial_state}. 

\begin{equation}
    \ket{x} \frac{1}{\sqrt{|E|}}\sum_{\theta \in E} \ket{\theta}\ket{0}\ket{y}\left( \frac{\ket{0}+\ket{1}}{\sqrt{2}} \right)
\label{eq:q_ensemble_intial_state}
\end{equation}

Implementing a quantum equivalent of a classifier such as a neural network requires a sequence of unitary operators of different types that combined form the quantum operator that implements it, such as in \cite{fawaz2019training}.
To simplify our examples,
let $\mathcal{F}$ be a unitary operator that implements the function $f(x, \theta)$, which is the function that returns the label given by the classifier, or the classifier's answer to input $x$.
Applying $\mathcal{F}$ to the system, the result would be as in equation \eqref{eq:q_ensemble_applying_f}.

\begin{equation}
    \begin{array}{c}
    \mathcal{F}\ket{x} \frac{1}{\sqrt{|E|}}\sum_{\theta \in E} \ket{\theta}\ket{0}\ket{y}\left( \frac{\ket{0}+\ket{1}}{\sqrt{2}} \right) \\
\downarrow \\
\ket{x} \frac{1}{\sqrt{|E|}}\sum_{\theta \in E} \ket{\theta}\ket{\hat{y}_\theta }\ket{y}\left( \frac{\ket{0}+\ket{1}}{\sqrt{2}} \right) \\
    \end{array}
\label{eq:q_ensemble_applying_f}
\end{equation}

The application of $\mathcal{F}$ would have to be repeated for every $x \in \mathcal{D}$. 
Where, everytime $\ket{\hat{y}_\theta} = \ket{y}$ the ancilary qubit would be rotated towards $\ket{0}$.
Thus, the degree of influence of each classifier would be decoded on the phase of the quantum state, as in \eqref{eq:q_ensemble_degrees_of_influence}.

\begin{equation}
    \frac{1}{\sqrt{|E|}}\sum_{\theta\in E} \ket{\theta}\left( \sqrt{a_\theta }\ket{0}+\sqrt{1 - a_\theta}\ket{1} \right) 
\label{eq:q_ensemble_degrees_of_influence}
\end{equation}

Afterward, the ancillary \textit{qubit} has to be measured. 
If after the measurement the \textit{qubit} assumes the state $\ket{1}$ the whole process of building the ensemble has to be made again. 
However, if after the measurement the \textit{qubit} assumes the state $\ket{0}$ the ensemble is ready to classify data samples from the validation set or any other unseen data.
The resulting state of the ensemble would be as in \eqref{eq:q_ensemble_final_state}.

\begin{equation}
\sum_{\theta \in E} \frac{\sqrt{a_\theta}}{\sqrt{\mathcal{X}|E|}} \ket{\theta}
\label{eq:q_ensemble_final_state}
\end{equation}

Where $\mathcal{X}$ is a normalization factor resulting from the measurement of the ancillary \textit{qubit}.
Let $\tilde{x}$ be an unseen datasample. 
In order to obtain the ensemble's answer, one would only need to add the data sample into the state, apply the $\mathcal{F}$ operator and measure the answer qubit in \eqref{eq:q_ensemble_aswer}.

\begin{equation}
\mathcal{F}\ket{\tilde{x}}\sum_{\theta \in E} \frac{\sqrt{a_\theta}}{\sqrt{\mathcal{X}|E|}} \ket{\theta}\ket{0} 
\rightarrow
\ket{\tilde{x}}\sum_{\theta \in E} \frac{\sqrt{a_\theta}}{\sqrt{\mathcal{X}|E|}} \ket{\theta}\ket{\hat{y}_\theta}
\label{eq:q_ensemble_aswer}
\end{equation}

However, by measuring the answer qubit the superposition of the ensemble would collapse, so the whole process of building the quantum ensemble would have to be repeated.
%A possible solution to avoid this kind of problem could be to entangle the answer register with another quantum register and obtain the answer using quantum teleportation, or another property from quantum physics, but this still requires further investigation.

\subsection{Replicating results from numerical analysis}
In \cite{schuld2018quantum} the authors analyze the use of an exponentially large quantum ensemble using quantum classifiers.
Presenting the idea of using such a quantum ensemble inspired in Bayesian learning, proposing a quantum classification method to be used as optimization free (or untrained) type of learning.
With the premiss that a reasonably large ensemble with weak but accurate classifiers would still return good classifications~\cite{kearns1988learning, schapire1990strength}.

A numerical analysis is presented further in \cite{schuld2018quantum} showing how such kind of ensemble would behave. 
The analysis was presented using simplified toy examples in a classical computer. 
The classes used in the examples were $Y = \{-1, 1\}$.

Where two binary classification problems were used, with the first case being an uni-dimensional case, where the function $f$ is to perform mappings of one-dimensional data points from both classes randomly generated with a normal distribution. 
Where the hyperparameters relative to each class will be referred to using a plus or minus signals, such as $\sigma_{+}$ and $\mu_{+}$ for the class $1$ and $\sigma_{-}$ and $\mu_{-}$ for class $-1$.

\begin{equation}
    f(x) = \frac{1}{\sigma_{\pm} \sqrt{2\pi}}e^{\left( -\frac{x-\mu_{\pm} }{\sqrt{2}\sigma_{\pm}} \right)^2}
\end{equation}

%Where in this case, the decision region of the classifier would be defined by calculating the mathematical expectancy of the classification according to the distribution of the data.
%It was possible to replicate some of their results. 

\begin{figure}[!ht]
    \centering
    \includegraphics[width=0.4\textwidth]{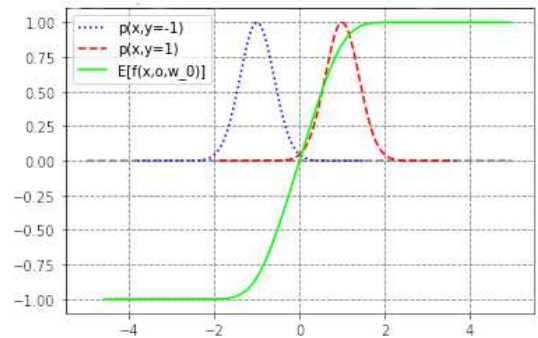}
    \caption{\label{fig:expectation_value} Decision region defined by expectatios value of the data points of the unidimentional case.}
\end{figure}

\begin{figure}[!ht]
    \centering
    \includegraphics[width=0.4\textwidth]{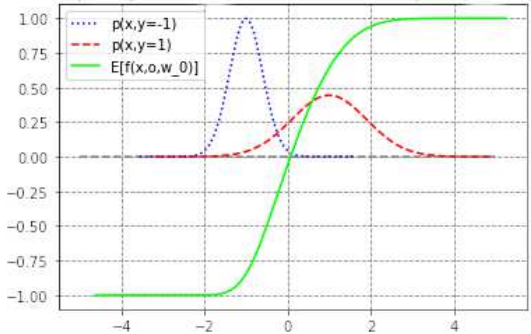}
    \caption{\label{fig:decision_boundary_shift} Shifting decision boundary by changing the distribtuions' variance of data points with class $1$ .}
\end{figure}

Fig. \ref{fig:expectation_value} shows some results of the replications concerning the calculation of the decision region for the one-dimensional case. 
The data points to be classified were randomly generated with the parameter $\sigma_{\pm} = 0.4$, and with $\mu_{-} = -1$ and $\mu_{+} = 1$. 
It can be seen that the decision boundary region can be found on the intersection of the distributions $p(\tilde{x}, y=-1)$ and $p(\tilde{x}, y=1)$ from data points with class $-1$ and $1$ respectively.
However, if the variance of one of the classes' distribution is changed, the decision boundary is shifted towards one of the classes' distribution as shown in Fig. \ref{fig:decision_boundary_shift}. 
Where the variance was changed from $\sigma_{+}=0.4$ to $\sigma_{+} = 0.9$.
Showing how this optimization free kind of classification is dependent on how well distributed are the data points for each class.

\begin{figure}[!ht]
    \centering
    \subfloat[]{\label{subfig:balanced_bidimensional_untrained}}{
    \includegraphics[width=0.4\textwidth]{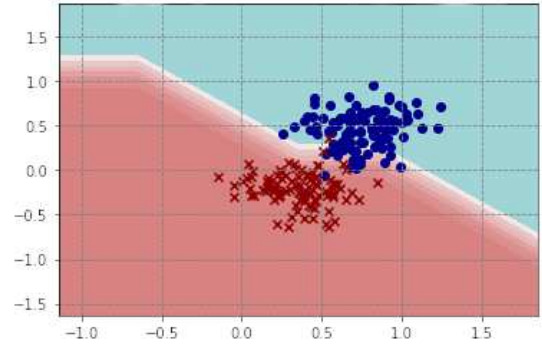}
    %\mbox{\epsfbox{../images/balanced_untrained_8000_2.eps}}
    }
    
    \subfloat[]{\label{subfig:balanced_bidimensional_trained}}{
    \includegraphics[width=0.4\textwidth]{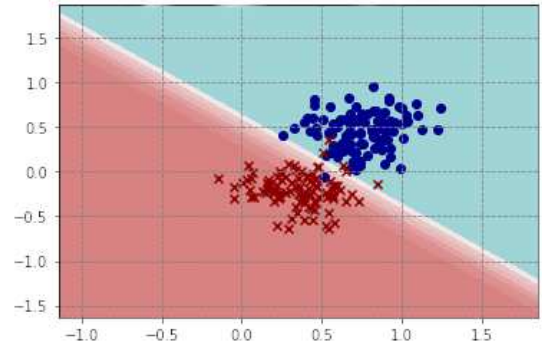}
    %\mbox{\epsfbox{../images/balanced_trained_8000_2.eps}}
    }
    \caption{\label{fig:balacend_bidimensional} Computing decision region for bi-dimensional data points.
    \ref{subfig:balanced_bidimensional_untrained} Decision boundary computed using $8000$ untrained perceptrons. 
    \ref{subfig:balanced_bidimensional_trained} Decision boundary computed using $8000$ trained perceptrons.
    }
\end{figure}

The second classification case was also a binary classification problem but, with bi-dimensional data points.
Where the data points used in the replication of this case were randomly generated using scikit-learn's~\cite{scikit-learn} $blob\_function$. 
With parameters used to generate the data points for each class being $\mu_{-}=[-1, 1]$ and $\mu_{+}=[1, -1]$.
With a standard deviation of $0.5$ for data points in both classes.
As in \cite{schuld2018quantum}, $8000$ perceptrons were created to classify the data points, with weights and biases in the interval $[-1, 1]$.
Fig. \ref{fig:balacend_bidimensional} shows the results of the replication of the bi-dimensional case for both trained and untrained perceptrons.
And as can be seen, the decision boundary finds in the center of both sets in the case using trained perceptrons.

\begin{figure}[!ht]
    \centering
    \subfloat[]{\label{subfig:unbalanced_bidimensional_untrained}}{
    \includegraphics[width=0.4\textwidth]{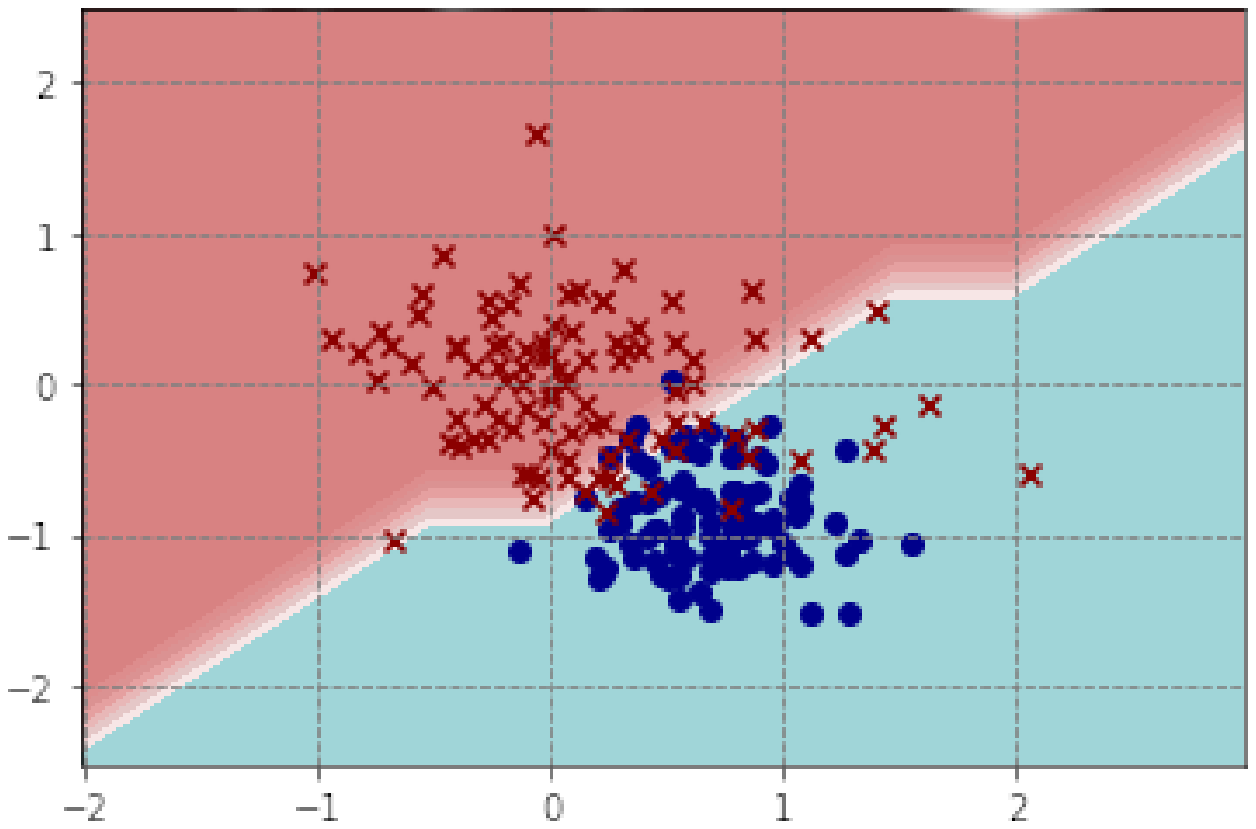}
    %\mbox{\epsfbox{../images/unbalanced_untrained_plus_8000.eps}}
    }
    
    \subfloat[]{\label{subfig:unbalanced_bidimensional_trained}}{
    \includegraphics[width=0.4\textwidth]{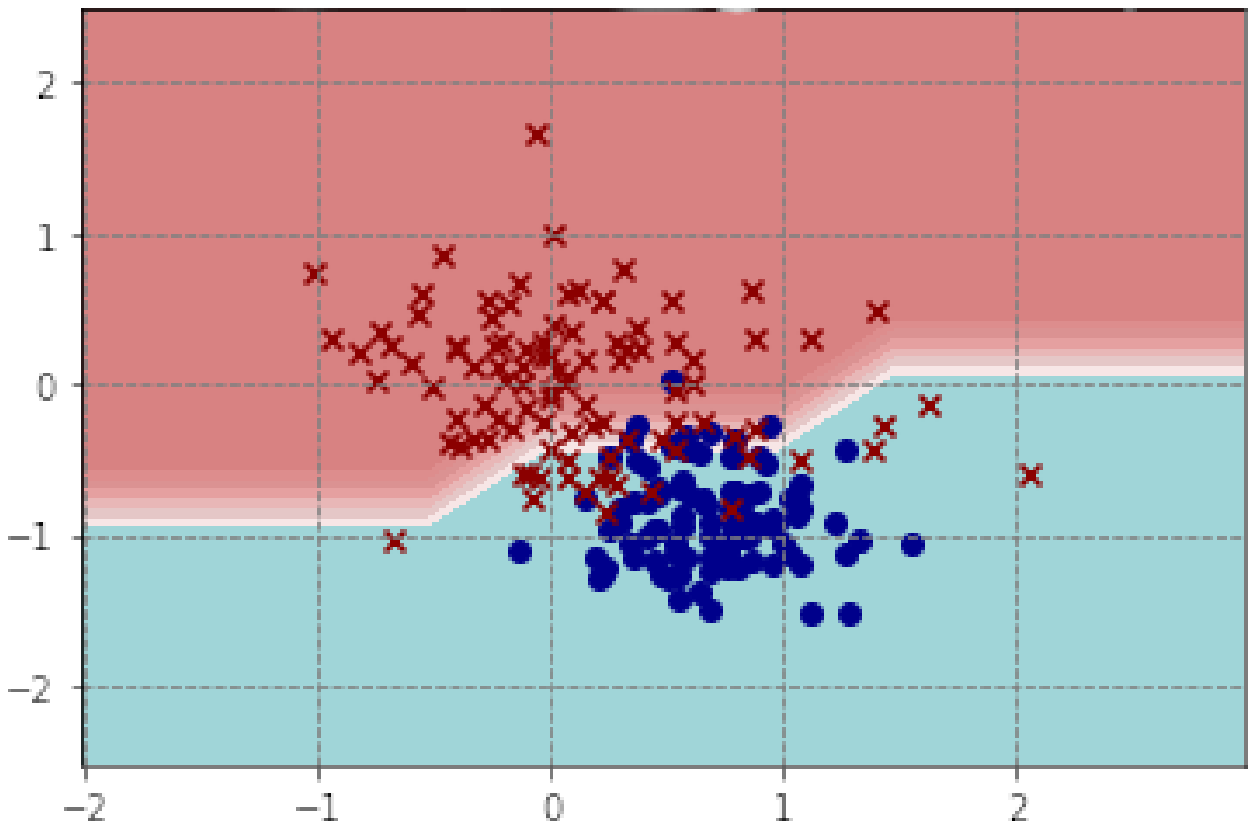}
    %\mbox{\epsfbox{../images/unbalanced_trained_plus_8000.eps}}
    }
    \caption{\label{fig:unbalacend_bidimensional} Computing decision region for bi-dimensional data points, with standard deviation of the distribution of data points with class $-1$ changed to $0.3$.
    \ref{subfig:balanced_bidimensional_untrained} Decision boundary computed using $8000$ untrained perceptrons. 
    \ref{subfig:balanced_bidimensional_trained} Decision boundary computed using $8000$ trained perceptrons.
    }
\end{figure}

We went a bit further with the replications and executed a bi-dimensional case, where the data points from both classes are distributed differently. 
Fig. \ref{fig:unbalacend_bidimensional} shows the new decision region dividing the two newly generated sets.
A difference in the definition of the decision boundary can be remarked between the cases using trained and untrained perceptrons.
Where the difference becomes less subtle when taking to account the outliers, or the points located further away from the center of each of their respective classes.

The replications presented in this work were made with the purpose to show how dependant an optimization free kind of ensemble would be to the data points distribution for each class.
However, those are but replications of toy examples of numerical analysis made in a classical manner, where the differences between trained and untrained models can be subtle.

\section{Quantum ensemble of trained classifiers}
\label{sec:methodology}
Taking into consideration how a quantum ensemble of quantum classifiers work.
We decided to evaluate what would be the results of such a quantum ensemble using benchmark datasets. 
The datasets used in this approach were the sets made available by the \textit{scikit-learn}'s~\cite{scikit-learn} library: iris, breast-cancer, and wine.

Unfortunately, quantum computers made available today are still very limited to perform this kind of experiment.
Therefore, we decided to make a simulation-based on calculations of the probability amplitudes in the system.
The type of classifier used in the simulations was the multilayer perceptron (MLP). 
All of them with 1 hidden layer with 10 neurons. 
The models were implemented using with the \textit{python} programming language, using the \textit{pytorch} library~\cite{paszke2017automatic}.

The cross-validation methodology used was hold-out. 
Being $70\%$ of the data points used for training and $30\%$ of the data points used for validation.
Whose accuracy values shall be referred to as training and validation respectively.

Two experiments were executed for the simulations and two metrics were used to evaluate both experiments.
The First metric was the Mean Probability Per Sample of the ensemble to return the correct label or $MPPS_{hit}$.
A quantum circuit returns its answer in a probabilistic way.
So, to compute the $MPPS_{hit}$ we first computed the probability of the system to return the correct label for each sample, a Probability Per Sample (or $PPS_{hit}$). 
Then, the arithmetic mean was calculated to obtain the $MPPS_{hit}$.

Another metric taken into account was the ensemble's overall accuracy, that is, the accuracy of the answers given by the set of classifiers. 
Let $p_x$ to be the probability of the ensemble to answer a datapoint correctly. 
In order to measure the ensemble's overall accuracy, a threshold probability ($p_t$) was used, so that when $p_x \geq p_t$ a correct answer would be computed to the ensemble's overall accuracy. 

The threshold probability value chosen for the experiments was $0.7$. 
This value was chosen after a series of trials and errors, in order to define a threshold that was not too low valued, lest bad performing ensemble would have a great influence on the final results.
And to define not a too much high valued threshold, lest it would be too prohibitive for reasonable performing ensembles to be computed in the final results.

During the execution of the experiments, the number of members in the ensemble was variated in the set $|E| \in \{100, 200, 300, 400, 500\}$.

\begin{algorithm}[!t]
\caption{Pseudocode of the simulations were performed}\label{alg:pseudocode_quantum}
    \begin{algorithmic}[1]
    \Procedure{quantum ensemble}{$\mathcal{D}$,$|E|$ }
        \For{ Each data set $D \in \mathcal{D}$}
            \State Create ensemble of classifiers $E = \{\theta_1, \ldots, \theta_{|E|}\}$
            \State Divide $D$ into training ($D_t$) and validation ($D_v$) sets
            \State Initialize the $array$ $PPS_{hit} \leftarrow \varnothing$
            \If{Use the training step}
                \For{Each training epoch $\mathcal{T}$}
                    \For{Each classifier $\theta_i \in \{\theta_1, \ldots, \theta_{|E|}\}$}
                        \State Train $\theta_i$                    
                        \State Calculate the accuracy $a_{\theta}$ using $D_t$
                    \EndFor
                \EndFor
               \Else                                  
                   \For{Each classifier $\theta_i \in \{\theta_1, \ldots, \theta_{|E|}\}$}                  
                       \State Calculate the accuracy $a_{\theta}$ using $D_t$
                  \EndFor
               \EndIf
        \State Create the $array$ of probability amplitudes $\left[\sqrt{ \frac{a_{\theta_1}}{ \mathcal{X} |E|} }, \ldots, \sqrt{ \frac{a_{\theta_{|E|}}}{ \mathcal{X} |E|} } \right]$
            \State Process $D_v$ with classifiers $\{\theta_1, \ldots, \theta_{|E|}\}$
            \For{Each datapoint $(x, y) \in D_v$}                
                \State Calculate $p_{hit} = \sum_{\theta \in E} \sqrt{ \frac{a_{\theta_{|E|}}}{ \mathcal{X} |E|}} $ for each classifier $\theta \in E$ that hits $y$       
                \If{$p_{hit} >= p_t$}
                    \State Increment overall accuracy of the ensemble
                \EndIf
                \State Save $p_{hit}$ into the $array$ $PPS_{hit}$
            \EndFor
            \State Compute the $MPPS_{hit}$ using the data from $array$ $PPS_{hit}$
        \EndFor            
    \EndProcedure
    \end{algorithmic}
\end{algorithm}

%\section{Results}
%\label{sec:results}

From the experiments executed, the first was to perform classification using untrained classifiers to verify the results of an optimization free learning over the benchmark datasets.
And the second kind of experiment was to perform classification using trained models to verify if by adding a training step in the system it is possible to improve the ensemble's performance, or if it remains unchanged.
Algorithm \ref{alg:pseudocode_quantum} contains pseudocode with a high-level explanation of how the experiments were executed.

%Results for untrained classifiers
\subsection{Simulating a quantum ensemble with untrained classifiers}
\begin{figure}[!ht]
	
	\centering
	\includegraphics[width=.35\textwidth]{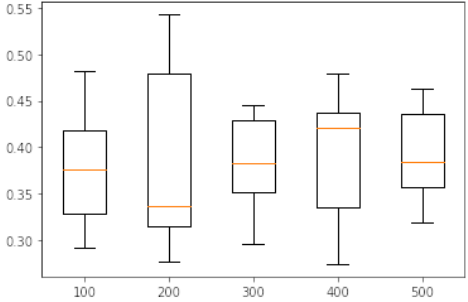}
	%\mbox{\epsfbox{../images/box_ploting_iris_dset_untrained_updated.eps}}
	\caption{\label{fig:untrained_iris} 
	        $PPS_{hit}$ of the ensembles with untrained classifiers on the Iris Dataset. 
			Where the $x$ axis show the number of members in the ensembles,
			and the $y$ axis are the range of propability values}
\end{figure}

In the simulations, the overall accuracy of a quantum ensemble using untrained classifiers was $0.0$ for all ensemble sizes defined, using a threshold of $0.7$.
Table \ref{tab:untrained_iris} shows the results concerning the $MPPS_{hit}$ and its standard deviation ($StdD$) from quantum ensembles of different sizes using untrained classifiers. 
It can be observed that for all ensemble sizes the $MPPS_{hit}$ remained below $0.5$ which explains the overall accuracy of $0.0$.
Fig. \ref{fig:untrained_iris} presents a sumarry of the $PPS_{hit}$ for an ensemble with different number of untrained classifiers.

\begin{table}[!ht]
    \caption{\label{tab:untrained_iris} Results from a quantum ensemble with untrained classifiers on the Iris dataset.}
    \centering
    \scalebox{.9}{
    \begin{tabular}{|c|c|c|}
    %\multicolumn{4}{c}{Iris} \\
    \hline
    $|E|$ & $MPPS_{hit}$ & $StdD$\\
    \hline
    $100$ & $0.37$   & $0.053$ \\
    \hline
    $200$ & $0.37$   & $0.084$ \\
    \hline
    $300$ & $0.38$   & $0.042$ \\    
    \hline
    $400$ & $0.39$   & $0.059$ \\
    \hline
    $500$ & $0.39$   & $0.043$ \\
    \hline
    \end{tabular}
    }
\end{table}

\begin{figure}[!t]
	\centering
	\includegraphics[width=.35\textwidth]{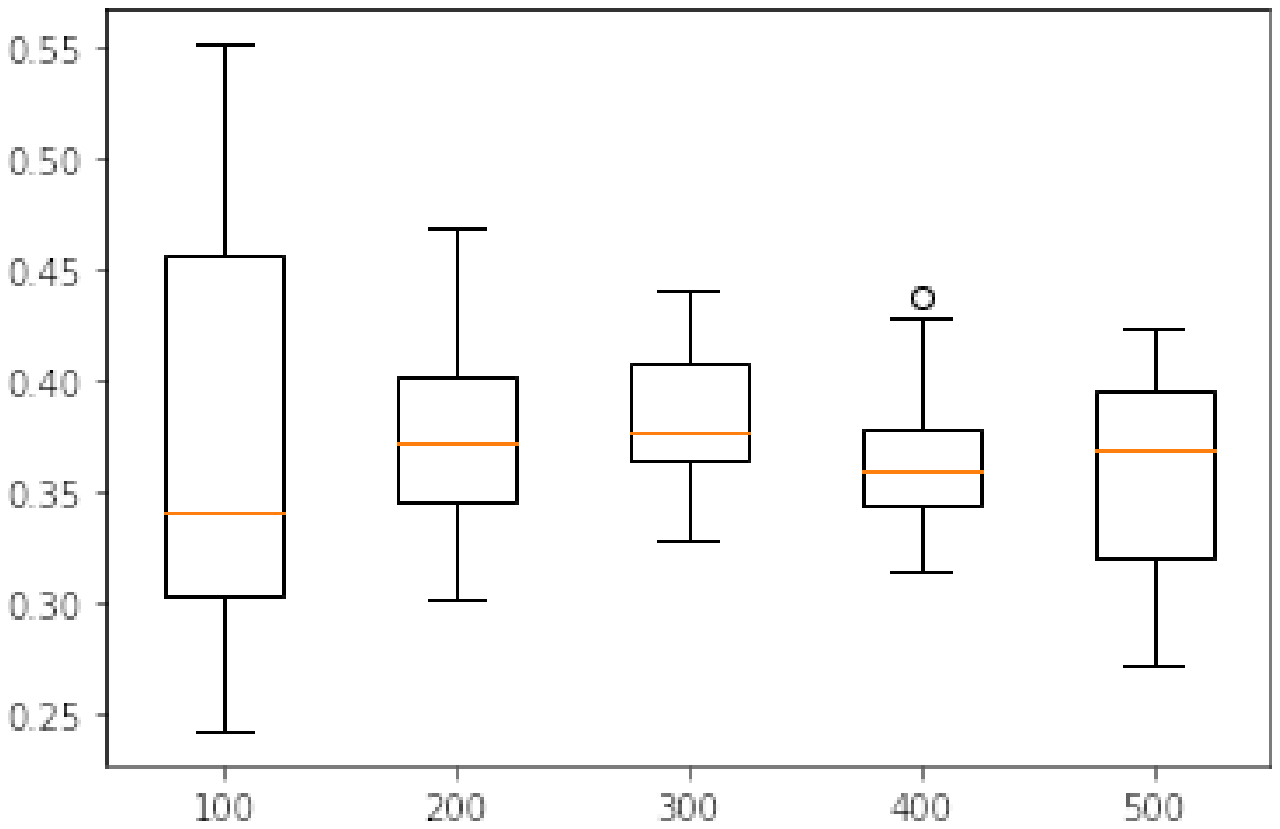}
	%\mbox{\epsfbox{../images/box_ploting_iris_dset_untrained_updated.eps}}
	\caption{\label{fig:untrained_wine}
			$PPS_{hit}$ of the ensembles with untrained classifiers on the Wine Dataset. 
			Where the $x$ axis show the number of members in the ensembles,
			and the $y$ axis are the range of propability values}
\end{figure}

\begin{figure}[!t]
	\centering
	\includegraphics[width=.35\textwidth]{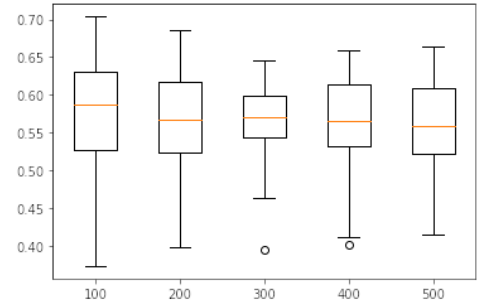}
	%\mbox{\epsfbox{../images/box_ploting_breast_cancer_dset_untrained.eps}}
	\caption{\label{fig:untrained_breast_cancer}
			$PPS_{hit}$ of the ensembles with untrained classifiers on the Breast Cancer Dataset. 
			Where the $x$ axis show the number of members in the ensembles,
			and the $y$ axis are the range of propability values}
\end{figure}

Tables \ref{tab:untrained_wine} and \ref{tab:untrained_breast_cancer} show the results concerning $MPPS_{hit}$ and its $StdD$ on the Wine and Breast-Cancer dataset.
From those results, slightly better performance of the quantum ensemble with untrained classifiers on the Breast-Cancer dataset can be observed relative to the Iris dataset, according to the values of $MPPS_{hit}$.
And slightly worse results are obtained with untrained classifiers on the Wine dataset. 
The results from both datasets toghether with the Iris dataset's results show that the overall $MPPS_{hit}$ on all the datasets remained below the threshold value of $0.7$, thus resulting in an overall accuracy of $0.0$ for all ensemble sizes on all the datasets.
Fig. \ref{fig:untrained_wine} and \ref{fig:untrained_breast_cancer}  presents a sumarry of the $PPS_{hit}$ for an ensemble with different number of untrained classifiers for both datasets.

\begin{table}[!ht]
\caption{\label{tab:untrained_wine} Results from a quantum ensemble with untrained classifiers on the Wine dataset.}    
    \centering
    \scalebox{.9}{
    \begin{tabular}{|c|c|c|}
    %\multicolumn{4}{c}{Wine} \\
    \hline
    $|E|$ & $MPPS_{hit}$ & $StdD$\\
    \hline
    $100$ & $0.37$       & $0.086$ \\
    \hline
    $200$ & $0.37$       & $0.04$ \\
    \hline
    $300$ & $0.38$       & $0.03$ \\    
    \hline
    $400$ & $0.36$       & $0.028$ \\
    \hline
    $500$ & $0.35$       & $0.041$ \\
    \hline
    \end{tabular}
    }
\end{table}

\begin{table}[!ht]
    \caption{\label{tab:untrained_breast_cancer} Results from a quantum ensemble with untrained classifiers on the Breast-Cancer dataset.}
    \centering
    \scalebox{.9}{
    \begin{tabular}{|c|c|c|}
    %\multicolumn{4}{c}{Breast-Cancer} \\
    \hline
    $|E|$ & $MPPS_{hit}$ & $StdD$ \\
    \hline
    $100$ & $0.54$       & $0.034$ \\
    \hline
    $200$ & $0.55$       & $0.038$ \\
    \hline
    $300$ & $0.54$       & $0.036$ \\    
    \hline
    $400$ & $0.52$       & $0.047$ \\
    \hline
    $500$ & $0.53$       & $0.029$ \\
    \hline
    \end{tabular}
    }
\end{table}

These results show that the optimization free approach is not a good option for a quantum ensemble when using real-life datasets.
To improve the ensemble's performance we decided to add a training (optimization) step, right after the creation of the models, as it is shown in Algorithm \ref{alg:pseudocode_quantum}.

%Results for trained classifiers -----------------------------------------------------------------------
\subsection{Simulating a quantum ensemble with trained classifiers}

In the simulations the models were trained with different training epochs, variating in the set $\{5, 10, 15\}$. 
In all training epochs for all values in $|E|$, the optimization method used in the training step as Stochastic Gradient Descent (SGD)~\cite{bottou2010large}.
With a learning rate of $1.2$ and a momentum of $0.9$ and a batch size of $10$ datapoints.
The values concerning the number of training epochs, batch size, learning rate, and momentum values were defined after a series of tests that involved trial and error. 
Tables \ref{tab:trained_iris}, \ref{tab:trained_wine} and \ref{tab:trained_breast_cancer} show the overall accuracy for every ensemble size in $|E|$ according to each of the tree training epochs number aforementioned.
As in the untrained case, the results are according to the Iris, Wine and Breast-Cancer datasets respectively.

\begin{figure}[!t]
	\centering
	\includegraphics[width=.35\textwidth]{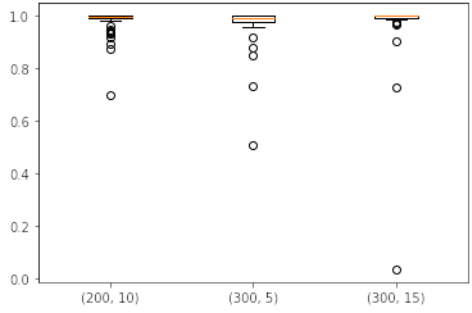}
	%\mbox{\epsfbox{../images/box_ploting_iris_dset_trained.eps}}
	\caption{\label{fig:trained_iris}
			$PPS_{hit}$ of the ensembles with trained classifiers on the Iris Dataset. 
			Where the $x$ axis show the number of members in the ensembles with  the respective training epochs,
			and the $y$ axis are the range of propability values}
\end{figure}

The best results for the Iris dataset in Table \ref{tab:trained_iris} were the quantum ensemble with $200$ and  $300$ ensemble members, both having an overall accuracy of $0.97$.
With the classifiers in the ensemble of size $200$ trained with $10$ training epochs.
Where the $MPPS_{hit}$ in this case was $0.97$, with $StdD$ of $0.051$.
And the classifiers trained in the ensemble of size $300$ trained with $5$ and $15$ training epochs.
Which for those cases the $MPPS_{hit}$ were both $0.96$, with $StdD$ of $0.083$ and $ 	0.146$ respective to each case of training epochs presented.
Figure \ref{fig:trained_iris} presents a summary of the $PPS_{hit}$ for the best results on the Iris data set.

%% IRIS
\begin{table}[!ht]
\caption{Overall accuracies of a quantum ensemble with different number of ensemble members over the \textbf{Iris} dataset}
    \label{tab:trained_iris}
    \centering
\begin{tabular}{|c|c|c|}
    %\multicolumn{3}{c}{\textit{iris}} \\
    \hline
    $|E|$                     & No. of epochs                 & overall accuracy\\
    \hline
    \multirow{3}{*}{$100$}   &        $5$                     & $0.93$         \\
                             \cline{2-3}                                           
                             &        $10$                    & $0.95$         \\
                             \cline{2-3}
                             &        $15$                    & $0.95$          \\
       \hline
    \multirow{3}{*}{$200$}   &        $5$                     & $0.95$         \\
                             \cline{2-3}
                             &        $10$                    & $0.97$         \\
                             \cline{2-3}
                             &        $15$                    & $0.93$         \\
    \hline
    \multirow{3}{*}{$300$}   &        $5$                     & $0.97$         \\
                             \cline{2-3}
                             &        $10$                    & $0.95$         \\
                               \cline{2-3}                            
                             &        $15$                    & $0.97$         \\                             
    \hline
    \multirow{3}{*}{$400$}   &        $5$                     & $0.88$         \\
                             \cline{2-3}                
                             &        $10$                    & $0.95$          \\
                             \cline{2-3}                
                             &        $15$                    & $0.95$         \\
    \hline
    \multirow{3}{*}{$500$}   &        $5$                     & $0.93$         \\
                             \cline{2-3}                
                             &        $10$                    & $0.95$         \\
                             \cline{2-3}                                          
                             &        $15$                    & $0.93$         \\
    \hline    
    \end{tabular}
\end{table}

The best results for the Wine dataset in Table \ref{tab:trained_wine} were from the quantum ensemble with $100$, $300$ and $400$ ensemble members, all three ensemble sizes containing results with an overall accuracy of $1.0$.
With the ensemble size of $100$ and $300$ with models trained with $15$ epochs, and the ensemble size of $400$ with models trained with $10$ epochs.
The $MPPS_{hit}$ for the quantum ensemble with $100$ and $300$ models was $0.98$, with a $StdD$ of $0.042$ and $0.032$ respectively.

\begin{figure}[!t]
	\centering
	\includegraphics[width=.35\textwidth]{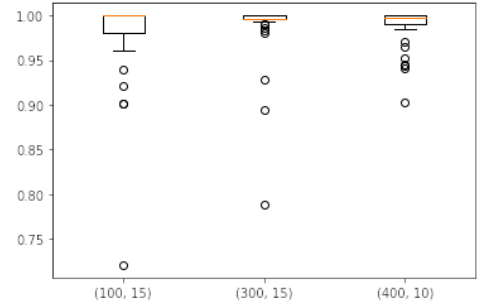}
	%\mbox{\epsfbox{../images/box_ploting_wine_dset_trained.eps}}
	\caption{\label{fig:trained_wine}
			$PPS_{hit}$ of the ensembles with trained classifiers on the Wine Dataset. 
			Where the $x$ axis show the number of members in the ensembles with  the respective training epochs,
			and the $y$ axis are the range of propability values}
\end{figure}

And $MPPS_{hit}$ for the quantum ensemble with $400$ models was $0.98$, with a $StdD$ of $0.018$. 
Figure \ref{fig:trained_wine} presents a summary of the $PPS_{hit}$ for the best results on the Wine data set.

%% WINE 
\begin{table}[!ht]
    \caption{Overall accuracies of a quantum ensemble with different number of ensemble members over the \textbf{Wine} dataset}
    \label{tab:trained_wine}
    \centering
    \begin{tabular}{|c|c|c|}
    %\multicolumn{3}{c}{Wine} \\
    \hline
    $|E|$                     & No. of epochs               & overall accuracy\\
    \hline
    \multirow{3}{*}{$100$}   &        $5$                   & $0.96$ \\
                             \cline{2-3}
                             &        $10$                  & $0.96$ \\                                            
                             \cline{2-3}
                             &        $15$                  & $1.00$  \\
                             
    \hline
    \multirow{3}{*}{$200$}   &        $5$                   & $0.98$  \\
                             \cline{2-3}
                             &        $10$                  & $0.96$  \\
                             \cline{2-3}
                             &        $15$                  & $0.96$ \\
    \hline
    \multirow{3}{*}{$300$}   &        $5$                   & $0.96$\\
                             \cline{2-3}    
                             &        $10$                  & $0.98$ \\
                               \cline{2-3}
                             &        $15$                  & $1.00$  \\                             
    \hline
    \multirow{3}{*}{$400$}   &        $5$                   & $0.96$\\
                             \cline{2-3}
                             &        $10$                  & $1.00$ \\
                             \cline{2-3}
                             &        $15$                  & $0.96$ \\
    \hline
    \multirow{3}{*}{$500$}   &        $5$                   & $0.98$  \\
                             \cline{2-3}
                             &        $10$                  & $0.92$ \\
                             \cline{2-3}
                             &        $15$                  & $0.98$ \\
    \hline    
    \end{tabular}
\end{table}

\begin{figure}[!t]
	\centering
	\includegraphics[width=.35\textwidth]{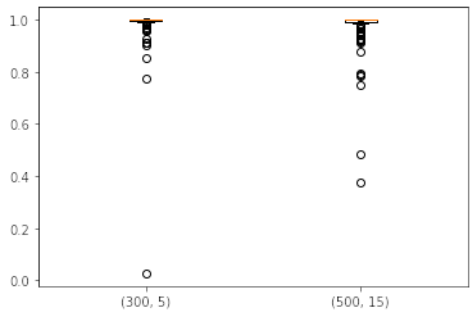}
	%\mbox{\epsfbox{../images/box_ploting_breast_cancer_dset_trained.eps}}
	\caption{\label{fig:trained_breast_cancer}
			$PPS_{hit}$ of the ensembles with trained classifiers on the Breast Cancer Dataset. 
			Where the $x$ axis show the number of members in the ensembles with  the respective training epochs,
			and the $y$ axis are the range of propability values}
\end{figure}

The best results for the Breast Cancer dataset in Table \ref{tab:trained_breast_cancer} were from quantum ensembles with $300$ and $500$ ensemble members, trained with $5$ and $15$ training epochs respectively. 
Where the $MPPS_{hit}$ for the quantum ensemble with $300$ members of $0.98$, with $StdD$ of $0.077$.
And the $MPPS_{hit}$ for the quantum ensemble with $500$ members of $0.98$, with $StdD$ of $0.07$. 
Figure \ref{fig:trained_breast_cancer} presents a summary of the $PPS_{hit}$ for the best results on the Breast Cancer data set.

%% BREAST-CANCER
\begin{table}[!ht]
    \caption{Overall Accuracies of a Quantum Ensemble With Different Number of Ensemble Members Over the \textbf{Breast-Cancer} Dataset}
    \label{tab:trained_breast_cancer}
    \centering
    \begin{tabular}{|c|c|c|}
    %\multicolumn{3}{c}{Breast-Cancer} \\
    \hline
    $|E|$                    & No. of epochs                & overall accuracy\\
    \hline
    \multirow{3}{*}{$100$}   &        $5$                   & $0.95$ \\
                             \cline{2-3}
                             &        $10$                  & $0.97$ \\                                                     \cline{2-3}
                             &        $15$                  & $0.95$  \\
                             
    \hline
    \multirow{3}{*}{$200$}   &        $5$                   & $0.95$      \\
                             \cline{2-3}
                             &        $10$                  & $0.95$  \\
                             \cline{2-3}
                             &        $15$                  & $0.95$ \\
    \hline
    \multirow{3}{*}{$300$}   &        $5$                   & $0.99$\\
                             \cline{2-3}
                             &        $10$                  & $0.95$ \\
                               \cline{2-3}
                             &        $15$                  & $0.95$  \\
                             
    \hline
    \multirow{3}{*}{$400$}   &        $5$                   & $0.97$\\
                             \cline{2-3}
                             &        $10$                  & $0.95$ \\
                             \cline{2-3}
                             &        $15$                  & $0.96$ \\
    \hline
    \multirow{3}{*}{$500$}   &        $5$                   & $0.97$  \\
                             \cline{2-3}
                             &        $10$                  & $0.95$ \\
                             \cline{2-3}
                             &        $15$                  & $0.98$ \\
    \hline    
    \end{tabular}
\end{table}

\section{Conclusions}
\label{sec:conclusions}
From the results presented we can infer that using a quantum ensemble of quantum classifiers as a form of optimization free learning is not a good approach to perform classification in a quantum computer.
Mainly because by simply adding a training step it was possible to significantly improve the performance of the ensemble relative to the overall accuracy and the mean probability of obtaining the correct answer. 

Because of the hardware limitations concerning the recent quantum computers, it was not possible to perform an experiment by using an exponentially large quantum ensemble, as it would be theoretically possible on an actual general-purpose quantum computer. 

Despite the fact the results in this work were obtained from simulations by calculating the probability amplitudes of the system, they serve as experimental evidence of the great advantage of still using a training step, or optimization step,  in a quantum ensemble of quantum classifiers, which is the main contribution of this work.  
Even if adding the optimization step means increasing the cost relative to gate application in building the quantum circuit.

The models used in the simulations were artificial neural networks.
Other works investigating the codification and training of neural networks can be found in literature~\cite{ricks2004training,da2016quantum,fawaz2019training}. 
%Therefore, the idea of training a neural network in a quantum system is not too distant from what is theoretically possible.
This work investigates the training of a quantum ensemble of quantum classifiers using neural networks with a fixed architecture. 
A future work might be the addition of architecture selection of neural networks~\cite{dos2018quantum2}.

Some quantum processors were made available, which can be accessed through the cloud. 
%Which is the case of the $IBM$ framework $Qiskit$~\cite{Qiskit}.
However, Noisy Intermediate Scale Quantum processors are still limited in the number of qubits. 
Another possible future work might be to make a reduced case scenario to be used in those currently small scale quantum processors.

\section*{Acknowledgement}
This work was supported by CNPq, CAPES and FACEPE (Brazilian research agencies).
% End Sections ----------------------------------------------------------------------

\bibliographystyle{bib_files/IEEEtran}
\bibliography{bibliography}

\end{document}